# Defect engineering and effect of vacancy concentration on the electrochemical performance of V-based MXenes

*Leiqiang Qin,[a,]\* Rutuparna Samal,[a] Jianxia Jiang,[a,b] Joseph Halim,[a] Ningjun Chen,[a] Florian Chabanais,[a] Per O. A. Persson,[a] Johanna Rosen[a,]\**

[a.]Department of Physics, Chemistry, and Biology (IFM), Linköping University, 58183 Linköping, Sweden

[b.]Flexible Electronics Innovation Institute (FEII), Jiangxi Provincial Key Laboratory of Flexible Electronics, Jiangxi Science and Technology Normal University, Nanchang 330013, China

ABSTRACT.

Vacancies play a pivotal role in determining the physical and chemical properties of materials. Introducing vacancies into two-dimensional (2D) materials offers a promising strategy for developing high-performance electrode materials for electrochemical energy storage. Herein, a facile top-down strategy is employed to create V-based MXenes with tunable vacancy concentrations, achieved by designing the precursor $(V_{1-x}Cr_x)_2AlC$ (x=0.05, 0.1, 0.3) MAX phase and precisely controlling the etching process. Systematic investigations reveal that introducing a moderate concentration of Cr-induced vacancies significantly enhances both the capacitance and rate performance of V-based MXenes. Specifically, $V_{1.9}CT_z$ achieves a capacitance of 760 F $g^{-1}$, far exceeding the 420 F $g^{-1}$ of vacancy-free $V_2CT_z$ MXene. In contrast, an excessively high vacancy concentration lead to deteriorated electrochemical performance and compromised structural stability. This work illustrates that defect engineering is a powerful approach to tailor the electrochemical properties of MXenes, offering a framework for designing next-generation MXene-based energy storage systems.

**Introduction**

The widespread use of fossil fuels has led to severe environmental pollution and an escalating energy crisis, thereby prompting the pursuit of sustainable and greener energy storage solutions.[1-3] In this context, electrochemical energy storage has attracted significant attention. However, the performance, cost, and reliability of energy storage devices are largely restricted by electrode materials. Thus, the rational design of high-performance electrode materials is both scientifically and practically important. Electrode materials that enable enhanced energy storage properties generally feature high surface area, increased porosity, excellent electrical conductivity, chemical stability, and abundant electrochemically active sites. In this context, metal oxides/mixed metal oxides[4,5], metal sulfides[6,7], MXenes[8-10], conducting polymers[11,12], and carbon-based materials[13,14] have been extensively explored. Despite these efforts, many electrode materials still suffer from slow reaction kinetics and sluggish ion transport, which ultimately compromise their electrochemical performance. As the internal electronic structure and composition of electrode materials govern these kinetics and transport processes, fine-tuning their structure becomes critical to achieving improved performance. Among various approaches, defect engineering has recently emerged as a powerful strategy to modify the surface properties and electronic structure of electrode materials, particularly at the nanoscale, thereby enhancing their electrochemical behavior.[15-20]

MXenes, a rapidly expanding family of two-dimensional transition metal carbides, nitrides, and carbonitrides, possess tunable surface chemistry, high electrical conductivity, redox-active surface sites, and hydrophilic surfaces.[21,22] These properties make them highly promising for applications such as energy storage[23-26], wireless communication[27], functional textiles[28,29], and bioelectronics[30,31]. Given their potential for energy storage, large efforts have been made to regulate their electrochemical properties. Among them, surface functionalization (utilizing terminations, $T_z$), intercalation, and defect engineering by chemical methodologies play an important role in fine-tuning the electrochemical properties.[32-34] In recent years, we have successfully created in-plane ordered vacancies in MXenes using a top-down manner. This is achieved by strategically incorporating reactive transition metals at the M sites within MAX phase precursors and subsequently removing specific elements from both M and A sites through precise chemical etching with F-containing acids. To facilitate the introduction of ordered vacancies, we realized a new family of quaternary in-plane ordered MAX (*i*-MAX) phases with the general formula $(M'_{2/3}M''_{1/3})_2AC$, in which M″=Sc, Y.[35,36] This unique structure allows for the selective removal of both Al and M″ elements during the etching process, resulting in in-plane ordered M″ vacancies within the delaminated M′$_{4/3}$CT$_z$ $i$-MXene nanosheets. For example, the Mo$_{4/3}$CT$_z$ $i$-MXene derived from a (Mo$_{2/3}$Sc$_{1/3}$)$_2$AlC 3D precursor exhibits excellent capacity performance (>1100 F cm$^{-3}$) and significantly improved conductivity compared to Mo$_2$CT$_z$[37], attributed to the introduced vacancies and related change in surface chemistry.[38] Furthermore, this defect engineering approach has been extended to other vacancy-containing MXene candidates, such as $i$-MXene W$_{4/3}$CT$_z$[39], which demonstrates outstanding electrocatalytic performance, as well as disordered-vacancy MXenes, Nb$_{1.33}$CT$_z$ with high electrochromic performance[40] and Mo$_{1.74}$CT$_z$ with improved capacity performance[41]. Based on these previous studies, it has been demonstrated that the introduction of vacancies can effectively adjust the physical and chemical properties of MXene materials. However, the intricate relationship between vacancy concentration and distribution with MXenes' electrochemical performance and stability remains underexplored. Therefore, developing simple and intuitive methods for vacancy optimization is crucial and holds the potential to drive significant breakthroughs in energy storage technologies.

Herein, we start from the 3D precursor design of vanadium (V)-based MAX phases, with a goal to precisely introduce metal vacancies of varying concentrations, in a top-down manner, in V-based MXenes. Utilizing a high-temperature solid-state reaction and adjusting the chromium (Cr) content, we synthesized a series of 3D (V$_{1-x}$Cr$_x$)$_2$AlC (x=0.05, 0.1, 0.3) quaternary MAX phase precursors, followed by hydrofluoric acid (HF) etching to selectively and simultaneously dissolve aluminum (Al) and Cr atoms, leading to the formation of MXene sheets with vacancies and pores dispersed throughout the structure. By varying the Cr content (x), we effectively tuned the vacancy concentration within the MXenes. Systematic investigation of the electrochemical properties of these vacancy-containing MXenes revealed their high capacity (760 F g$^{−1}$), excellent rate capability, and good stability at low vacancy concentrations (V$_{1.9}$C). Our work demonstrates that regulating vacancy concentration through defect engineering significantly enhances the electrochemical properties, providing new insights for the design and application of functional MXenes.

**Results and Discussion**

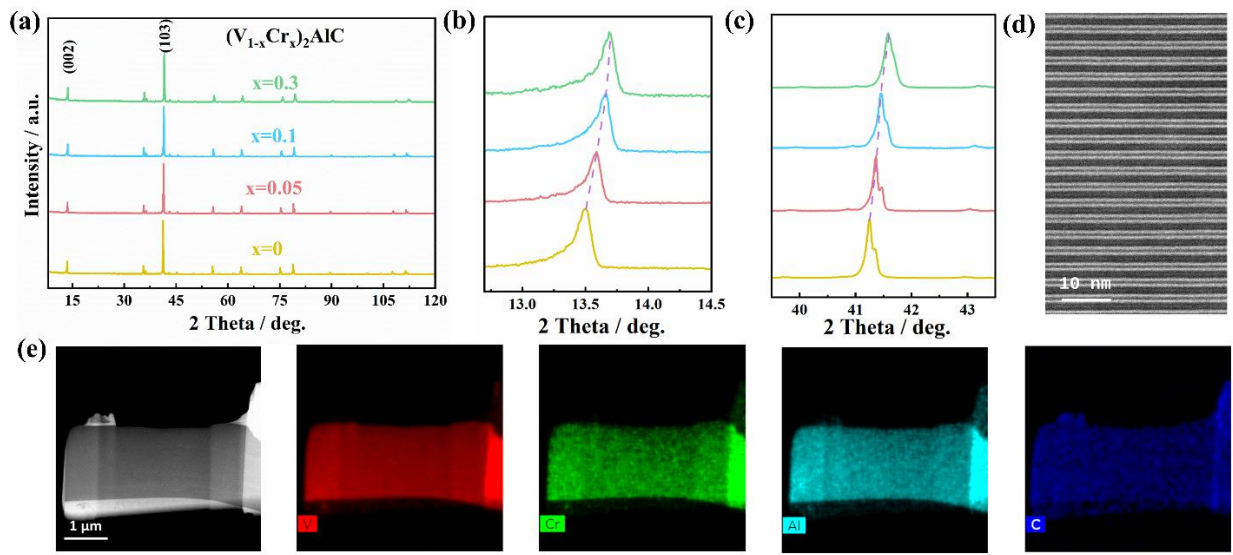

**Figure 1.** (a) X-ray diffraction pattern of the $(V_{1-x}Cr_x)_2AlC$ (x=0, 0.05, 0.1, 0.3) MAX phase. (b, c) The magnified XRD patterns of $(V_{1-x}Cr_x)_2AlC$. (d) STEM image of $(V_{0.95}Cr_{0.05})_2AlC$ MAX phase. (e) STEM-EDX elemental mapping images of the $(V_{0.95}Cr_{0.05})_2AlC$ MAX phase.

A series of 3D $(V_{1-x}Cr_x)_2AlC$ (x=0, 0.05, 0.1, 0.3) quaternary solid solution MAX phases were synthesized through a high-temperature solid-state reaction approach (Figure S1). From the XRD patterns of the prepared $(V_{1-x}Cr_x)_2AlC$ MAX phases, the absence of any peak broadening and splitting indicates that the chemical compositions of Cr and V are uniform in the $(V_{1-x}Cr_x)_2AlC$ MAX phases and that a solid solution is formed (**Figure 1a**). With increasing Cr content, the (002) and (103) diffraction peak of $(V_{1-x}Cr_x)_2AlC$ progressively shifts to higher angle (Figure 1b and c), consistent with a contraction of the $c$ lattice parameter caused by the smaller atomic radius of the Cr (139 pm) compared to V (153 pm). As the Cr content in the MAX phase increases, the average lattice parameter *c* gradually decreases, indicating that the solid solution MAX phase was successfully prepared. The XRD patterns of the prepared $(V_{1-x}Cr_x)_2AlC$ MAX phases show high purity, potentially ensuring the high quality of the generated MXene. Moreover, the high purity of the samples is also supported by SEM and energy-dispersive X-ray spectroscopy (EDX) analysis (Figure S2), where the measured V:Cr:Al ratio very closely match the nominal compositions of 2-2x:2x:1. A representative STEM image of the $V_{1.9}Cr_{0.1}AlC$ MAX phase (Figure 1d) reveals the typical layered structure of the MAX phase, and the low magnification EDX elemental maps (Figure 1e) confirm the homogeneous distribution of V, Cr, and Al. These results collectively demonstrate the formation of a uniform solid solution MAX phase suitable for subsequent MXene production.

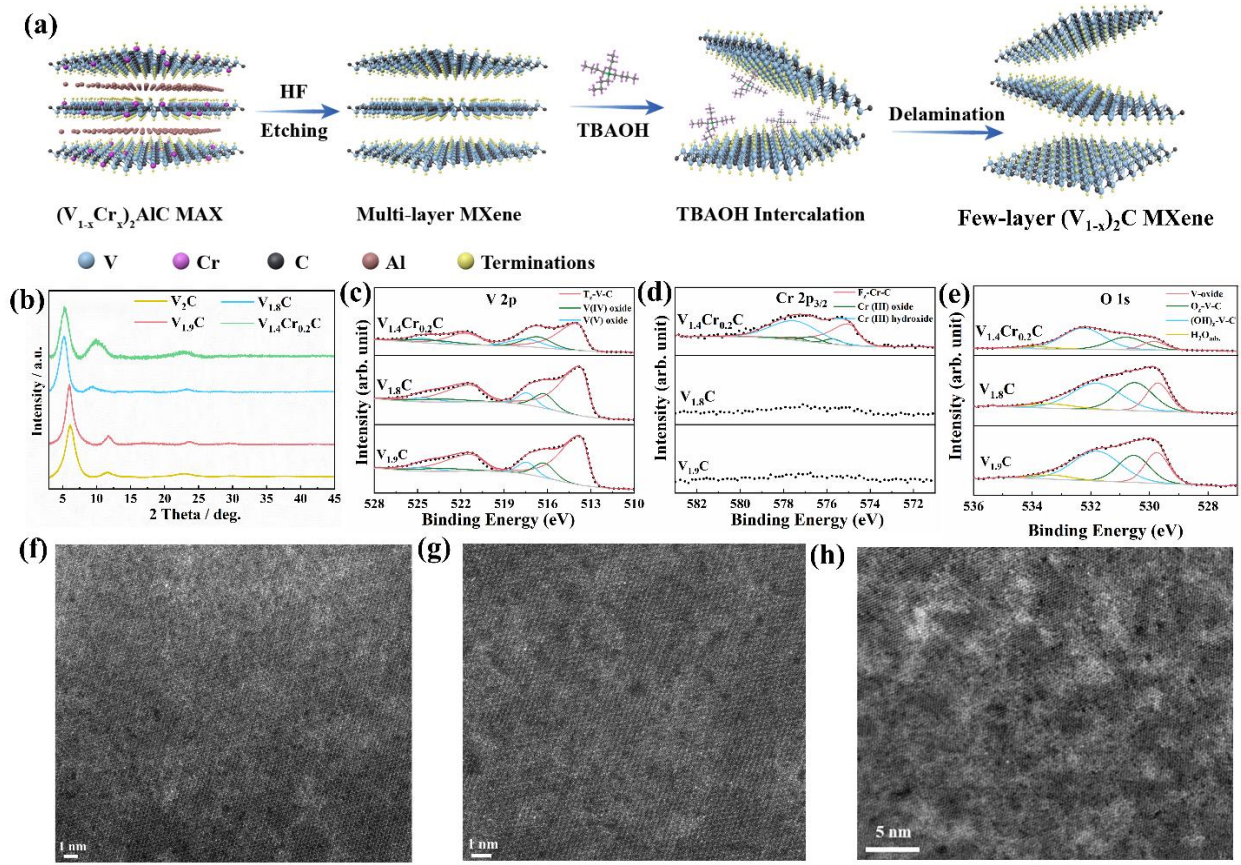

**Figure 2.** (a) Schematics of a $(V_{1-x}Cr_x)_2AlC$ MAX phase, its etching, and MXene delamination route. (b) X-ray diffraction pattern of $(V_{1-x})_2C$ MXene films. XPS spectra of $(V_{1-x})_2C$ (x = 0.05, 0.1, 0.3) MXene free-standing films for the regions (c) V 2p, (d) Cr $2p_{3/2}$ and (e) O 1s. (f-h) STEM images of $(V_{1-x})_2C$ MXene, x=0.05, 0.1, 0.3, respectively.

The multi-step protocol for converting a $(V_{1-x}Cr_x)_2AlC$ MAX phase into a vacancy-rich $(V_{1-x})_2C$ MXene is illustrated in **Figure 2a**. Notably, from this point forward we omit the general notation $T_z$ for the MXene terminations. In this top-down approach, Al and Cr are selectively removed via wet-chemical etching, yielding free-standing "paper" films through vacuum filtration (Figure S3). The disappearance of the characteristic MAX phase peaks in the XRD patterns (Figure 2b), accompanied by the emergence of a broad peak at low angles, confirms the successful preparation of $(V_{1-x})_2C$ MXenes. Notably, the introduced Cr render milder etching conditions (Table S1), presumably because the surface-exposed Cr preferentially dissolves, enhancing the penetration and transport of the etchant. Specifically, the MAX phase containing chromium can be etched at a lower temperature, a shorter time and a lower HF concentration than $V_2AlC$, and the yield does not decrease except for x=0.3, see details of the etching conditions in Table S1. All $(V_{1-x})_2C$ MXenes exhibit a broader (002) peak that shifts to lower angles as the Cr content increase, reflecting changes in interlayer spacing. X-ray photoelectron spectroscopy (XPS) and electron energy loss spectroscopy (EELS) were employed to characterize the compositions and chemical states of these MXenes (Figure 2c-e, Figure S4-6 and Table S2-8). In the high-resolution V 2p spectra (XPS, Figure 2c), the dominant signal arises from V-C bonds in the MXene structures, with minor $V^{4+}$ and $V^{5+}$ species assigned to vanadium surface oxides which might have formed during etching or subjecting the sample to the ambient. The XPS spectrum of the Cr 2p region (Figure 2d) confirms that only the $V_{1.4}Cr_{0.6}AlC$ MAX phase has Cr remaining after etching. The presence of Cr-C bonds confirms partial substitution of V by Cr, whereas Cr-O signals stem from oxygen-containing surface terminations. Based on the XPS results, the obtained vacancy-engineered MXene can be denoted as $V_{1.9}C$, $V_{1.8}C$, and $V_{1.4}Cr_{0.2}C$. At the same time, due to the presence of Cr in $V_{1.4}Cr_{0.2}C$, the binding energy of O-metal in the MXene moves toward the higher energy direction, indicating that there is a strong interaction between the oxygen-containing termination and the transition metal (V and Cr). High-resolution STEM images (figure 2f-h, Figure S7-9) reveal that each $(V_{1-x})_2C$ MXene nanosheet hosts abundant in-plane vacancies and pores. As the Cr content increases, the lateral size of the nanosheets decreases, while the density of visible pores increases, consistent with dynamic light scattering (DLS) measurements (Figure S10). These observations highlight that systematic modulation of Cr content in the parent MAX phases enables precise control over the vacancy concentration in V-based MXenes.

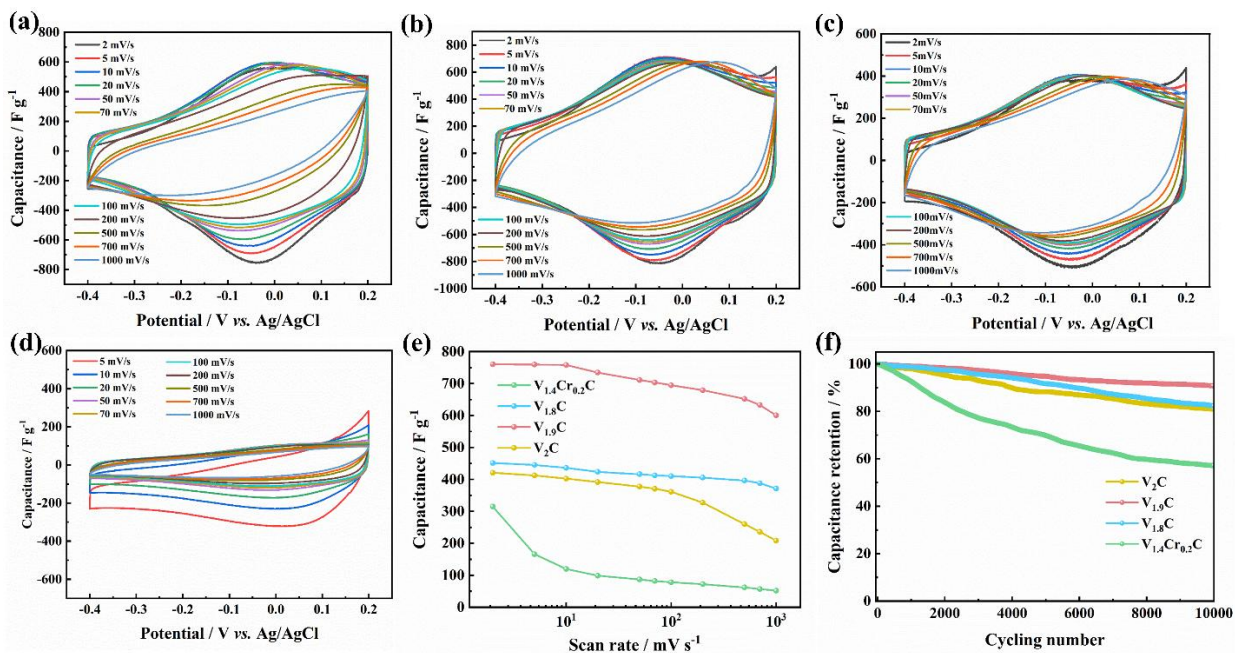

**Figure 3.** Electrochemical performance of $(V_{1-x})_2C$ MXene in 3 M $H_2SO_4$ in a three-electrode configuration. CV curves of (a) $V_2C$ MXene, (b) $V_{1.9}C$ MXene, (c) $V_{1.8}C$ MXene, and (d)

$V_{1.4}Cr_{0.2}C$ MXene from 2 to 1000 mV s$^{-1}$; (e) Comparison of the rate performance. (f) Long-term cycling performance.

As mentioned above, systematically varying the Cr content in the MAX phase precursor $(V_{1-x}Cr_x)_2AlC$ enables the synthesis of $(V_{1-x})_2C$ MXene with increasing vacancy concentrations and pore sizes. Because metal vacancies can markedly alter both the microstructure and the electronic structure of MXenes, their influence on electrochemical performance was investigated by comparing $(V_{1-x})_2C$ MXenes, including $V_2C$, $V_{1.9}C$, $V_{1.8}C$, $V_{1.4}Cr_{0.2}C$, in a three-electrode system in 3 $_M$ $H_2SO_4$ solution. **Figure 3a-d** display the cyclic voltammetry (CV) curves of $(V_{1-x})_2C$ MXene at scan rates ranging from 2 mV s$^{-1}$ to 1000 mV s$^{-1}$ within a stable potential window of -0.4 to 0.2 V. All samples except $V_{1.4}Cr_{0.2}C$ display a pair of pronounced redox peaks between -0.1 and 0 V. Notably, $V_{1.9}C$ exhibits markedly enhanced electrochemical behavior compared to vacancy-free $V_2C$, with its specific capacitance reaching 760 F g$^{-1}$ at a scan rate of 2 mV s$^{-1}$ (the highest reported for a V-based MXene to date, Table S9) along with exception rate performance (Figure 3e): the capacitance decays only 21% following a scan rate increase from 2 to 1000 mV s$^{-1}$ (600 F g$^{-1}$). Furthermore, $V_{1.9}C$ exhibited superior cycling stability, maintaining a retention of 90.9% after 10,000 cycles (Figure 3f). The beneficial role of moderate vacancy concentrations is highlighted, which appears to increase the density of electroactive sites and facilitate efficient ion transport. Although $V_{1.8}C$ shows a modest improvement over $V_2C$ in capacity, its overall performance lags behind $V_{1.9}C$, particularly with respect to cycling stability. Interestingly, $V_{1.8}C$ demonstrates the best high-rate performance (17.7% decay from 2 to 1000 mV s$^{-1}$), suggesting that moderately higher vacancy levels may favor rapid charge-discharge kinetics. However, when the vacancy concentration becomes excessive, as in $V_{1.4}Cr_{0.2}C$, the CV peaks become distorted and eventually disappear, reflecting compromised structural integrity and poor electrochemical stability.

Excessive introduction of metal vacancies will lead to an increase in the number of defects in the MXene layers, increase in the number of holes in the nanosheets, and disruption of its structural continuity. This makes the material more brittle, leading to the formation of smaller nanosheets during the synthesis process (Figure 2a), which will reduce the conductivity and have a negative impact on the electrochemical kinetics and ion transport performance, while causing the collapse of the nanostructure during ion insertion and extraction. These observations collectively highlight the importance of tuning the vacancy concentration to maximize specific capacitance, rate capability, and long-term cycling performance in MXene electrodes.

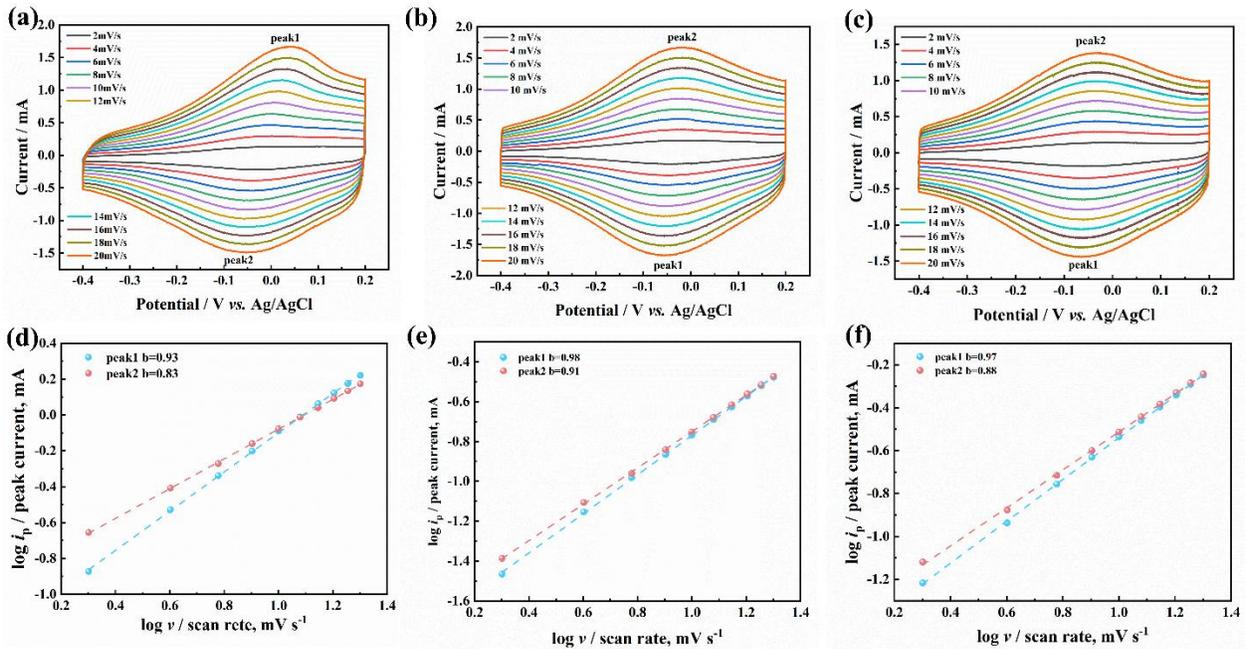

**Figure 4.** Electrochemical kinetics analysis of the $(V_{1-x})_2C$ MXenes. CV curves of (a) the $V_2C$ MXene, (b) the $V_{1.9}C$ MXene and (c) the $V_{1.8}C$ MXene between -0.4 and 0.2 V (*vs.* Ag/AgCl) at various scan rates from 2 to 20 mV s$^{-1}$. The corresponding log (peak current) versus log (scan rate) plots at the redox peak for (d) the $V_2C$ MXene, (e) the $V_{1.9}C$ MXene and (f) the $V_{1.8}C$ MXene.

To further investigate the impact of metal vacancy concentration on the charge storage mechanism of MXene, CV curves with scan rates ranging from 2 to 20 mV s$^{-1}$ were recorded (**Figure 4a-c**). The homologous processes of charge storage can be discerned by the linear relationship between log $i$ versus log $v$, using the following equation

$$i_p = av^b \quad (1)$$

where the peak current $i$ and sweep rate $v$ are observables and $a$, $b$ are adjustable parameters. The slope of the log($v$)-log($i$) plot can be employed to determine the value of $b$, which offers crucial information on the charge storage kinetics. In general, the value of $b$ is estimated to be between 1.0 and 0.5, indicating the capacitive process and diffusion-controlled process, respectively. For V$_2$C (Figure 4a,d), V$_{1.9}$C (Figure 4b,e), V$_{1.8}$C (Figure 4c,f), the $b$ values were calculated to be 0.93, 0.98, 0.97, respectively, for the cathodic peak, and 0.83, 0.91, and 0.88, respectively, for the anodic peak, which shows that the electrochemical properties of (V$_{1-x}$)$_2$C MXene are mainly characterized by a pseudocapacitive behavior. Notably, V$_{1.9}$C demonstrates the highest $b$ values, underscoring the dominant contribution of capacitive currents and reflecting optimized ion transport pathways that arise from a carefully tuned vacancy concentration.

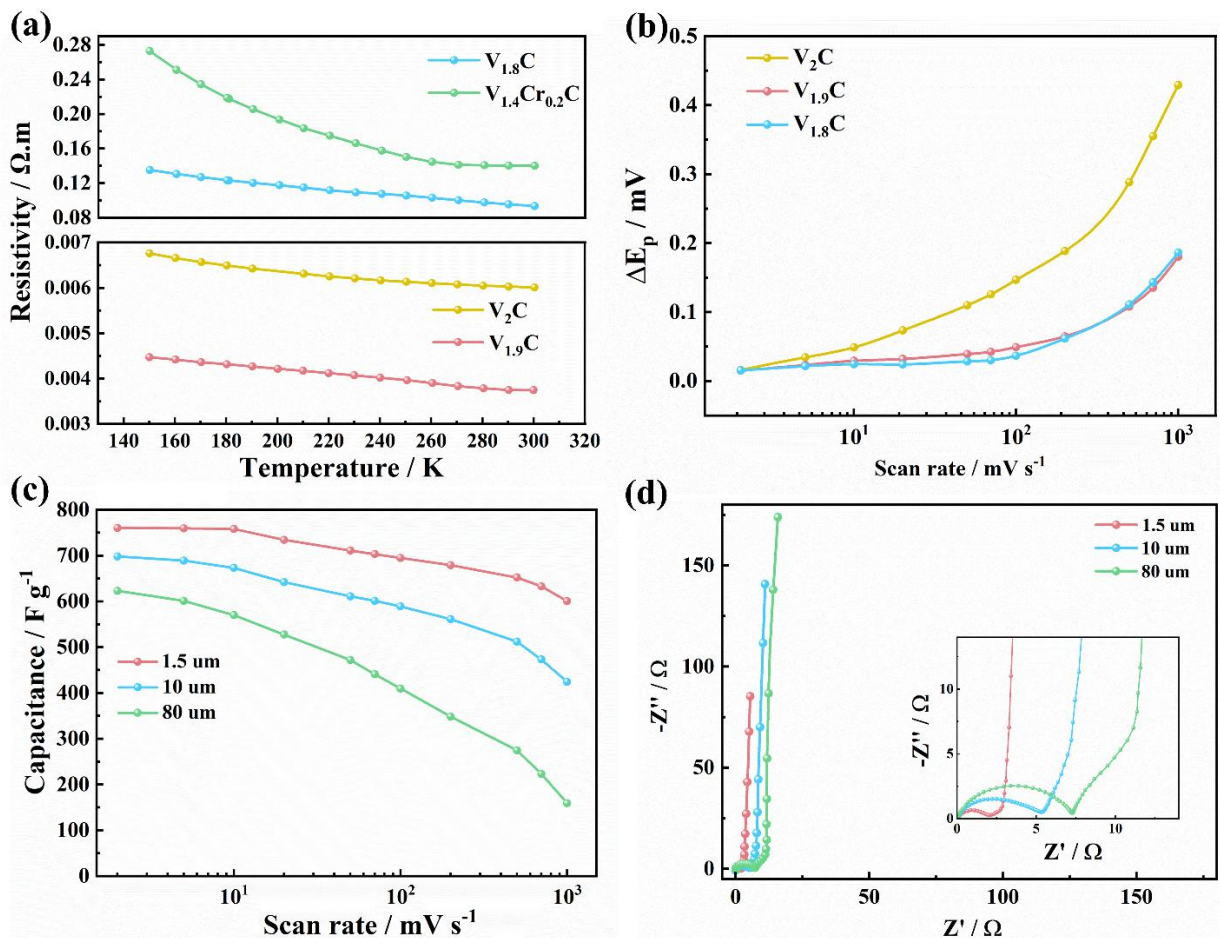

**Figure 5.** (a) Temperature-dependent resistivity of V-based MXenes. (b) Peak separation, $\Delta E_p$, for different scan rates extracted from CV analysis of $V_2C$, $V_{1.9}C$, and $V_{1.8}C$. (c) Rate performance of $V_{1.9}C$ films of different thickness. (d) Electrochemical impedance spectroscopy data collected from $V_{1.9}C$ films of different thickness.

The electronic and ion transport properties of vanadium-based MXenes with different vacancy concentration were further studied. The temperature-dependent electrical conductivity of the materials was investigated under vacuum conditions (**Figure 5a**). The results show that the resistance increases with increasing concentration of metal vacancies. Notably, $V_{1.9}C$ exhibits the lowest resistance of 0.0038 $\Omega$.m at 300 K, outperforming both vacancy-free $V_2C$ (0.006 $\Omega$.m) and samples with higher vacancy content, $V_{1.8}C$ (0.094 Ω.m) and $V_{1.4}Cr_{0.2}C$ (0.140 Ω.m). Moreover, $V_{1.9}C$ displays a near-constant resistance across the measured temperature range, suggesting excellent electron transport properties. These observations support the notion that optimally introduced vacancies can change the electronic state in a way that reduces resistivity[42], whereas excessive defect concentrations disrupt conduction pathways and degrade conductivity. In addition, defects also influence the electrochemical charge-storage processes, as indicated by comparing the cathodic and anodic peak separation ($\Delta E_p$) in the CV curves (Figure 5b). Specifically, both $V_{1.9}C$ and $V_{1.8}C$ exhibit smaller peak potential separation than $V_2C$, signifying enhanced redox reversibility-a result of the synergistic improvements in electron and ion transport afforded by a moderate vacancy density. In practical "paper" electrodes, horizontally stacked MXene flakes hinder ion transport and electron transfer at high thicknesses, causing diminished capacity and rate performance. However, for $V_{1.9}C$ with optimized vacancies, the specific capacitance drops by merely 10% (from 760 F g$^{-1}$ to 623 F g$^{-1}$) when the thickness is increased from 1.5 μm to 80 μm (Figure 5d). Even at 80 μm thickness, the capacity decays to 60% as the scan rate increases from 2 to 1000 mV s$^{-1}$. In contrast, $V_2C$ undergoes a more substantial decline, from 420 to 220 F g$^{-1}$, as the electrode thickness increases from 1.8 to 73 μm (Figure S11), accompanied by a pronounced decrease in rate capability (7.4% retention). The electrochemical impedance spectroscopy (EIS) data for $V_{1.9}C$ (Figure 5d) confirm its robust ion and electron transport even at elevated thicknesses, as evidenced by only a slight increase in the 45° Warburg-type region (related to ion transport resistance) and minimal alteration to the Nyquist plot slope, indicating that the introduction of defects reduces the sensitivity of the ion transport performance in MXene related to the film thickness. Overall, these findings demonstrate that precisely tuning the vacancy concentration, from a top-down perspective by introducing sacrificial metals in the MAX precursors, allows simultaneous optimization of both capacitance and ion transport in V-based MXenes. While shown here for vanadium-based systems, the same strategy can be broadly applicable to other MAX phase compositions, with the potential to unlock similarly superior or even improved performance. Moreover, the excellent ion transport in thick electrodes highlights a promising avenue toward scalable, high-performance MXene-based energy-storage architectures.

**Conclusion**

In this work, a top-down defect engineering approach was used to introduce controlled vacancy concentrations into V-based MXenes by rationally designing the precursor MAX phase, aiming to investigate the impact of vacancy concentration on the electrochemical properties. Specifically, Cr was incorporated at the M-site in $(V_{1-x}Cr_x)_2AlC$, generating a family of solid-solution MAX phases with systematically varied Cr content. Subsequent etching of these precursors simultaneously removed Al and Cr, producing V-based MXenes with different vacancy concentrations. The experimental results show that as the Cr content increases, that is, the vacancy concentration increases, this promotes the formation of additional in-plane pores, reduces lateral flake dimensions, and diminishes electrical conductivity. Notably, compared with $V_2AlC$, the introduction of chromium into the MAX phase can reduce the etching temperature, shorten the etching time and maintain high productivity. The sample with the lowest vacancy concentration, $V_{1.9}C$, achieved a markedly enhanced supercapacitance of 760 F $g^{-1}$, compared to 420 F $g^{-1}$ for vacancy-free $V_2C$ MXene. These findings underscore that an optimal level of vacancy incorporation can significantly boost the electrochemical performance, whereas excessive vacancies degrade both conductivity and structural integrity. Overall, this work highlights defect engineering as a robust strategy for tuning the properties of MXenes, complementing established strategies that modulate the X-site, layer number (n), and surface terminations ($T_z$). By expanding the palette of tunable parameters, this approach offers new opportunities for the rational design of MXenes across a broad range of advanced applications.

**Experimental Section**

*Synthesis of MAX phases.*

The bulk 3D $(V_{1-x}Cr_x)_2AlC$ MAX phase was prepared by a high-temperature solid state calcination method at a temperature of 1450 ˚C, with heating for three hours under constant Ar atmosphere (5 sccm) and a ramping rate of 10 ˚C min$^{-1}$. The detailed synthesis protocol involves pure elemental Vanadium powder (~325 mesh, 99.5%, Sigma Aldrich), Chromium powder (< 45 μm, ⩾ 99%), Aluminum powder (~325 mesh, 99.5%, Alfa Aesar), and graphite powder (~325 mesh, 99.9995%, Thermo scientific) as raw materials. These commercial single-element powders are mixed in stoichiometric molar ratios of 2(1-x):2x:1.2:1 (where x = 0, 0.05, 0.1, 0.3) with a total weight of 10 grams, then uniformly grinded using an agate mortar and pestle for 30 min. The blended precursor mixture was pre-compacted into pellets (dimension: thickness ~ 1 cm and diameter 0.5 cm) using a die set and a hydraulic press, applying a pressure of 5 tons for 60 s. The disk-shaped pellets were transferred into an alumina crucible with a lid and placed in the tubular furnace. After synthesis, the furnace was allowed to cool to room temperature automatically at a cooling rate of 5 ˚C min$^{-1}$. The sintered sample blocks were subsequently polished using sandpaper of 800 mesh to remove the outmost oxide layer and ground into a particle size of ~ 36 μm.

*Synthesis of MXene*

The $(V_{1-x}Cr_x)_2AlC$ MAX phase was used for derivation of MXene. In a typical procedure, 1 g of the MAX phase powder was added into 20 ml of hydrofluoric (HF) acid solution at a specific temperature under constant magnetic stirring of 500 rpm for a certain time (see details in Table S1). After the etching process was completed, the resulting solution was washed with nitrogenated deionized (DI) water and centrifuged at a rotation speed of 6000 rpm several times to remove the unwanted byproducts. Delamination was facilitated by the addition of 10 ml tetrabutylammonium hydroxide (TBAOH) solution, for enhanced interlayer spacing. The as-etched MXene solution was shaken manually for 10 min and washed three times with DI water, slowly rinsing the side walls and decanting the transparent solution. The diluted TBAOH-treated colloidal solution was centrifuged for 30 min at 3000 rpm. The high-quality delaminated suspension was decanted gently by excluding the sediment. In order to calculate the yield of synthesized MXene, first, 5 ml of the suspension was vacuum filtered to obtain free-standing film, which was then vacuum dried at room temperature overnight and weighed to calculate the concentration of the suspension.

*Characterization*

Crystallographic analysis of freshly synthesized 3D $(V_{1-x}Cr_x)_2AlC$ and etched MXene products was performed by using a Panalytical X'pert diffractometer operating at 40 kV and 40 mA with Cu $K_\alpha$ as radiation source ($\lambda$=1.5406 Å), with a step size of 0.0084 and 2θ ~ 3-120˚. The experimental configuration included divergence slits, receiving slits set at 1/2°, and a parallel plate collimator in conjunction with a Ni beta filter on the other side of the diffracted beam respectively. The optical beam setup for the incident beam consists of a parallel plate collimator with a Cu $K_\beta$ filter. Microstructural features and elemental composition were examined with scanning electron microscopy (SEM) using a GeminiSEM 560 Zeiss instrument equipped with an energy dispersive spectrometer (EDX). X-ray photoelectron spectroscopy (XPS) measurements were conducted on MXene free-standing films using Kratos AXIS Ultra$^{DLD}$ surface analysis system from Manchester, U.K., utilizing a monochromatic Al-Kα (1486.6 eV) radiation source. The X-ray beam irradiated the surface of the sample at an angle of 45º, with respect to the surface and provided an X-ray spot of 300 x 800 μm. Charge neutralization was performed using a co-axial, low energy (~0.1 eV) electron flood source to avoid binding energy (BE) shifts. XPS spectra were recorded for V 2p, Cr $2p_{3/2}$, C 1s, O 1s, F1s and Al 2p regions. The analyzer pass energy used for all the regions was 20 eV with a step size of 0.1 eV. The BE scale of all XPS spectra was referenced to the Fermi-edge ($E_F$), where $E_f = 0$ eV. The peak fitting was carried out using CasaXPS Version 2.3.16 RP 1.6 in the same manner as in Ref.[43] Transmission electron microscopy (TEM) images were acquired with the Linköping double-corrected FEI Titan[3] 60-300, operated at 300 kV. Electron energy loss spectroscopy (EELS) in both the low-loss and core-loss regions was facilitated by Dual-EELS. Electronic transport properties of MXenes after heat treatments were measured in a Quantum Design EverCool II Physical Property Measurement System (PPMS). Temperature-dependent resistance was recorded from 300 K down to 150 K in a low pressure helium environment (~20 Torr).

*Electrochemical characterization*

The filtered free-standing films were directly punched to 6 mm diameter and treated as working electrode without combining with carbon black and binder. The electrochemical assessments were recorded using a VMP3 potentiostat from Biologic, France. Polytetrafluoroethylene (PTFE) Swagelok cells were used for electrochemical measurements. The nanoporous polypropylene membrane from Celgard 3501, LLC (Limited liability company) with a specific pore size of 0.064 μm and YP-50 activated carbon (Kuraray, Japan) electrodes prepared using activated carbon with poly(vinylidene fluoride) is applied as a separator and counter electrode respectively. In three electrode experiments, Ag/AgCl in 1 M KCL as the reference electrode and 3 M $H_2SO_4$ solution served as the electrolyte. Cyclic voltammetry and galvanostatic charge-discharge tests were conducted at a fixed potential window. The electrochemical impedance spectroscopy (EIS) was executed at open circuit potential with an amplitude of 5 mV ranging in a frequency from 10 mHz to 100 kHz.

ASSOCIATED CONTENT

**Supporting Information**.

The Supporting Information is available free of charge.

AUTHOR INFORMATION

**Corresponding Authors**

Leiqiang Qin - Department of Physics, Chemistry, and Biology (IFM), Linköping University, 58183 Linköping, Sweden; Email: leiqiang.qin@liu.se

Johanna Rosen - Department of Physics, Chemistry, and Biology (IFM), Linköping University, 58183 Linköping, Sweden; Email: johanna.rosen@liu.se

**Author Contributions**

L.Q. and J.R. conceived and designed the research. L.Q. and R.S. performed the experiments and characterized the materials. J.J., J.H., N.C., F.C. and P.P. contributed to the completion of part of the characterizations. J.H., N.C., F.C. and P.P. analyzed all the data. All authors contributed to the discussion of the manuscript.

ACKNOWLEDGMENT

This work was supported by the Knut and Alice Wallenberg (KAW) foundation through a Scholar grant (2019.0433) and Project funding (2020.0033), the Swedish Government Strategic Research Area in Materials Science on Advanced Functional Materials at Linköping University,

Faculty Grant SFOMat-LiU 2009-00971, the European Union (ERC, MULTI2D, 101087713), the Technological Expertise and Academic Leaders Training Program of Jiangxi Province-Youth Program (20243BCE51085) and Jiangxi Provincial Natural Science Foundation (20224BAB214022). The authors acknowledge the Swedish Research Council (VR) and the Swedish Foundation for Strategic Research (SSF) for access to ARTEMI, the Swedish National Infrastructure in Advanced Electron Microscopy (nos. 2021-00171 and RIF21-0026).

(43) Naguib, M.; Halim, J.; Lu, J.; Cook, K. M.; Hultman, L.; Gogotsi, Y.; Barsoum, M. W. New two-dimensional niobium and vanadium carbides as promising materials for Li-ion batteries. *J. Am. Chem. Soc.* **2013**, *135*, 15966.25